# A 3-D Track-Finding Processor for the CMS Level-1 Muon Trigger

D.Acosta, B.Scurlock*, A.Madorsky, H.Stoeck, S.M.Wang

*Department of Physics, University of Florida, Gainesville, FL 32611, USA*

V.Golovtsov, M.Kan, L.Uvarov

*High Energy Physics Division, Petersburg Nuclear Physics Institute, Gatchina, Leningradskaya oblast, 188300, Russia*

*presented by Bobby Scurlock (bslock@phys.ufl.edu)

We report on the design and test results of a prototype processor for the CMS Level-1 trigger that performs 3-D track reconstruction and measurement from data recorded by the cathode strip chambers of the endcap muon system. The tracking algorithms are written in C++ using a class library we developed that facilitates automatic conversion to Verilog. The code is synthesized into firmware for field-programmable gate-arrays from the Xilinx Virtex-II series. A second-generation prototype has been developed and is currently under test. It performs regional track-finding in a 60 degree azimuthal sector and accepts 3 GB/s of input data synchronously with the 40 MHz beam crossing frequency. The latency of the track-finding algorithms is expected to be 250 ns, including geometrical alignment correction of incoming track segments and a final momentum assignment based on the muon trajectory in the non-uniform magnetic field in the CMS endcaps.

## 1. INTRODUCTION

The endcap regions of the Compact Muon Solenoid (CMS) experiment will consist of four stations of Cathode Strip Chambers (CSC). These chambers will provide CMS complete azimuth coverage (in $\phi$), as well as 0.9 to 2.4 in pseudo-rapidity ($\eta$). Six cathode strip chambers compose a single station in the endcap system. The chambers are trapezoidal in shape, extending $10^o$ or $20^o$ in $\phi$, and are composed of cathode strips aligned radially from the beam axis, and anode wires aligned in the orthogonal direction.

In the endcap system, muon track-finding is electronically partitioned into six $60^o$ sectors. A single Sector Processor (SP) receives trigger primitives from front-end electronics which sit on or near the CSCs. The front-end electronics form Local Charged Tracks (LCTs) from the six detector layers of a station. A Muon Port Card (MPC) collects the LCTs for a given station and sector, sorts them, and sends the best three to an SP via optical fibers. A single SP collects LCTs sent via fifteen 1.6 Gbit/s optical links, and is responsible for linking LCTs in $\phi$ and $\eta$ in order to form full tracks, and to report the transverse momentum ($p_t$), $\phi$, and $\eta$ for each track. The entire Track-Finding processor is composed of twelve such SPs housed in a single 9U VME crate. The challenge for the Track-Finding Processor is to report muon candidates with the lowest possible Pt threshold, and yet maintain a single muon trigger rate below 1 kHz/$\eta$ at full Large Hadron Collider (LHC) luminosity.

This paper is organized as follows: Section 2 will describe the Track-Finder (TF) algorithm, Section 3 will discuss the first prototype, Section 4 will discuss the second prototype including preliminary test results, Section 5 will discuss software used for the SP, and a summary can be found in Section 6.

## 2. TRACK-FINDER LOGIC

The principle of the TF logic [1] is illustrated in Figure 1. The Track-Finding process is partitioned into several steps. A given station within a sector may have as many as three LCTs reported to the SP. These LCTs are then converted into track-segments, which are described in terms of their $\phi$, and $\eta$ coordinates. Each track-segment in each station should be checked against the other segments in neighboring stations for consistency to share a single track. Thus, each track-segment is extrapolated through to other stations, and compared against existing segments. If an extrapolation is successful, these segments are "linked" to form a single track. Each possible pairwise combination is tested in parallel. After extrapolation, doublets are then linked to assemble full tracks. Redundant tracks are cancelled, the best three tracks are selected, and the track parameters are then measured.

The SP has the ability to handle LCTs received out of step from the actual bunch crossing time in which they originated. This is accomplished by the Bunch Crossing Analyzer, which allows the Sector Processor to form tracks from LCTs received up to one bunch crossing later than the earliest LCT.

The first step in the track-finding process is to extrapolate pairwise combinations of track-segments. This is accomplished by requiring the two segments to be consistent with a muon originating from the collision vertex and with an appropriate curvature induced by the non-uniform magnetic field. A successful extrapolation is assigned when two stubs lie within allowed windows of $\phi$ and $\eta$ - neither LCT should parallel to the beam axis, and both should appear to originate from the interaction region.

The Track Assembler Units (TAUs) examine successfully extrapolated track-segment pairs to see if a larger track can be formed. If so, those segments are combined and a code is assigned to denote which muon stations are involved.

A list of nine possible tracks is sent to the Final Selection Unit (FSU). Since different data streams may contain data from the same muons, the FSU must cancel redundant tracks, and select the best three distinct candidates.

The final stage of processing in the TF is the measurement of the track parameters, which includes the $\varphi$ and $\eta$ coordinates of the muon, the magnitude of the transverse momentum $P_T$, the sign of the muon, and an overall quality which we interpret as the uncertainty of the momentum measurement. The most important quantity to calculate accurately is the muon $P_T$, as this quantity has a direct impact on the trigger rate and on the efficiency. Simulations have shown that the resolution of the momentum measurement in the endcap using the displacement in $\varphi$ measured between two stations is about 30% at low momenta, when the first station is included. (It is worse than 70% without the first station.) We would like to improve this so as to have better control on the overall muon trigger rate, and the most promising technique is to use the $\varphi$ information from three stations when it is available. This improves the resolution to approximately 22% at low momenta, which is sufficient. In order to achieve a three-station $P_T$ measurement, we have developed a scheme that uses the minimum number of bits necessary in the calculation.

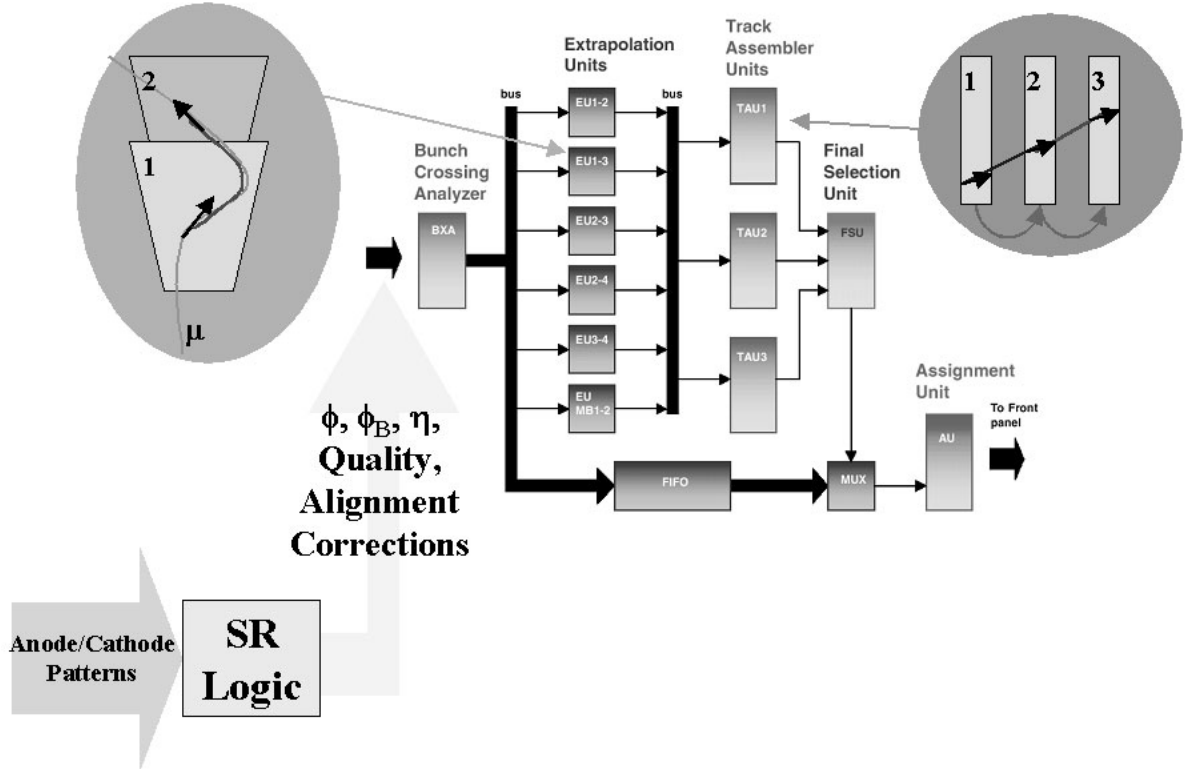

**Figure1: Sector Processor logic principle**

The first step is to do some pre-processing in FPGA logic: the difference in $\varphi$ is calculated between the first two track segments of the muon, and between the second and third track segments when they exist. Only the essential bits are kept from the subtraction. For example, we do not need the same accuracy on the second subtraction because we are only trying to untangle the multiple scattering effects at low momenta. The subtraction results are combined with the $\eta$ coordinate of the track and the track type, and then sent into a 4 MB memory for assignment of the signed $P_T$. Tracks composed of only two track segments are allowed also in certain cases.

## 3. FIRST PROTOTYPE SYSTEM ARCHITECTURE

In the first prototype [2], the Track-Finder is implemented as 12 Sector Processors that identify up to the three best muons in 60° azimuthal sectors. Each Processor is a 9U VME card housed in a crate in the counting house of CMS. Three Sector Receiver (SR) cards collect the optical signals from Muon Port Cards of that sector and transmit data to the Sector Processor via a custom point-to-point backplane. The SRs convert local CSC data into regional $\phi$ and $\eta$ coordinates to be used by the Track-Finder. A maximum of six track segments are sent from the first muon station in that sector, and three each from the remaining three stations. In addition, up to eight track segments from chambers at the ends of the barrel muon system are propagated to a transition board in the back of the crate and delivered to each Sector Processor as well.

A total of nearly 600 bits of information are delivered to each Sector Processor at the beam crossing frequency of 40 MHz (3 GB/s). To reduce the number of connections, LVDS Channel Link transmitters/receivers from National Semiconductor [3] are used to compress the data by about a factor of three through serialization/de-serialization. A custom point-to-point backplane operating at 280 MHz is used for passing data to Sector Processor.

Each Sector Processor measures the track parameters ($P_T$, $\varphi$, $\eta$, sign, and quality) of up to the three best muons and transmits 60 bits through a connector on the front panel. A sorting processor accepts the 36 muon candidates from the 12 Sector Processors and selects the best 4 for transmission to the Global Level-1 Trigger.

A prototype Sector Processor was built using 15 large Xilinx Virtex FPGAs, ranging from XCV50 to XCV400, to implement the track-finding algorithm, and one XCV50 for the VME interface (Figure 2).

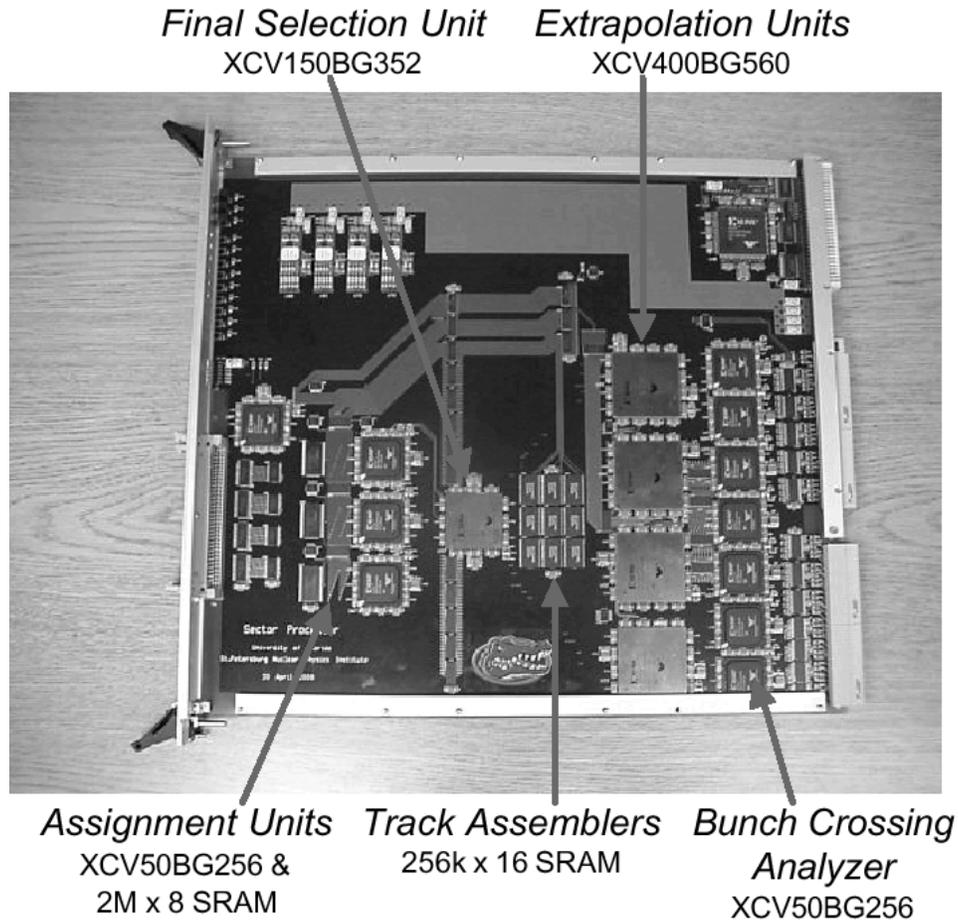

**Figure 2: First Sector Processor prototype**

The first prototype was completely debugged and tested. Simulated input data or random numbers were transmitted over the custom backplane to the SP, and the results were read from an output FIFO. These results were compared with a C++ model, and were shown to be in perfect agreement. The latency from the input of the Sector Receivers [4] (not including the optical link latency) to the output of the Sector Processor is 21 clocks, 15 of which are used by Sector Processor logic.

Computing in High Energy and Nuclear Physics, La Jolla, California, March 24-28, 2003            4

## 4. SECOND (PRE-PRODUCTION) PROTOTYPE SYSTEM ARCHITECTURE

Although the first prototype was proven a success, improvements were needed in the overall Level-1 Trigger design in order to meet the latency allowance of 3.2 µs. Recent significant improvements in logic density [5] allow implementing all Sector Processor logic onto one Xilinx Virtex-II FPGA (XC2V4000). Also, the optical link components have become smaller and faster; thus allowing the merger of three Sector Receivers and one Sector Processor of the first prototype onto one sixteen-layer board (Figure 3).

This board accepts fifteen optical links from the Muon Port Cards, where each link carries information

Because the track segment information arrives from 15 different optical links, it is aligned to the proper bunch crossing number.

The Sector Receiver algorithm is implemented in a series of cascaded look-up memories, using 45 GSI SRAMs [6], in order to minimize the size of SRAM chips required: chamber specific LCT data is sent to the first memory, this memory then sends a local ϕ measurement to two more memories which then use this local ϕ measurement along with the original LCT data to form a global ϕ and η measurement for the given sector (figure 4).

The angular information for all track segments is then passed to the main XC2V4000 FPGA, which executes the entire 3-dimentional tracking algorithm of the Sector corresponding to one muon track segment, which is described by a 32-bit word (two 16-bit frames). Each link transmits a 32-bit word every 25 ns. Finisar's FTRJ-8519-1-2.5 transceiver is used by the MPC for transmission, and by the SP for reception of LCT data. Additionally, the board receives up to 8 muon track segments, sent using LVDS synchronous with the 40 MHz LHC clock, from the Barrel Muon system via a transition card behind the custom backplane.

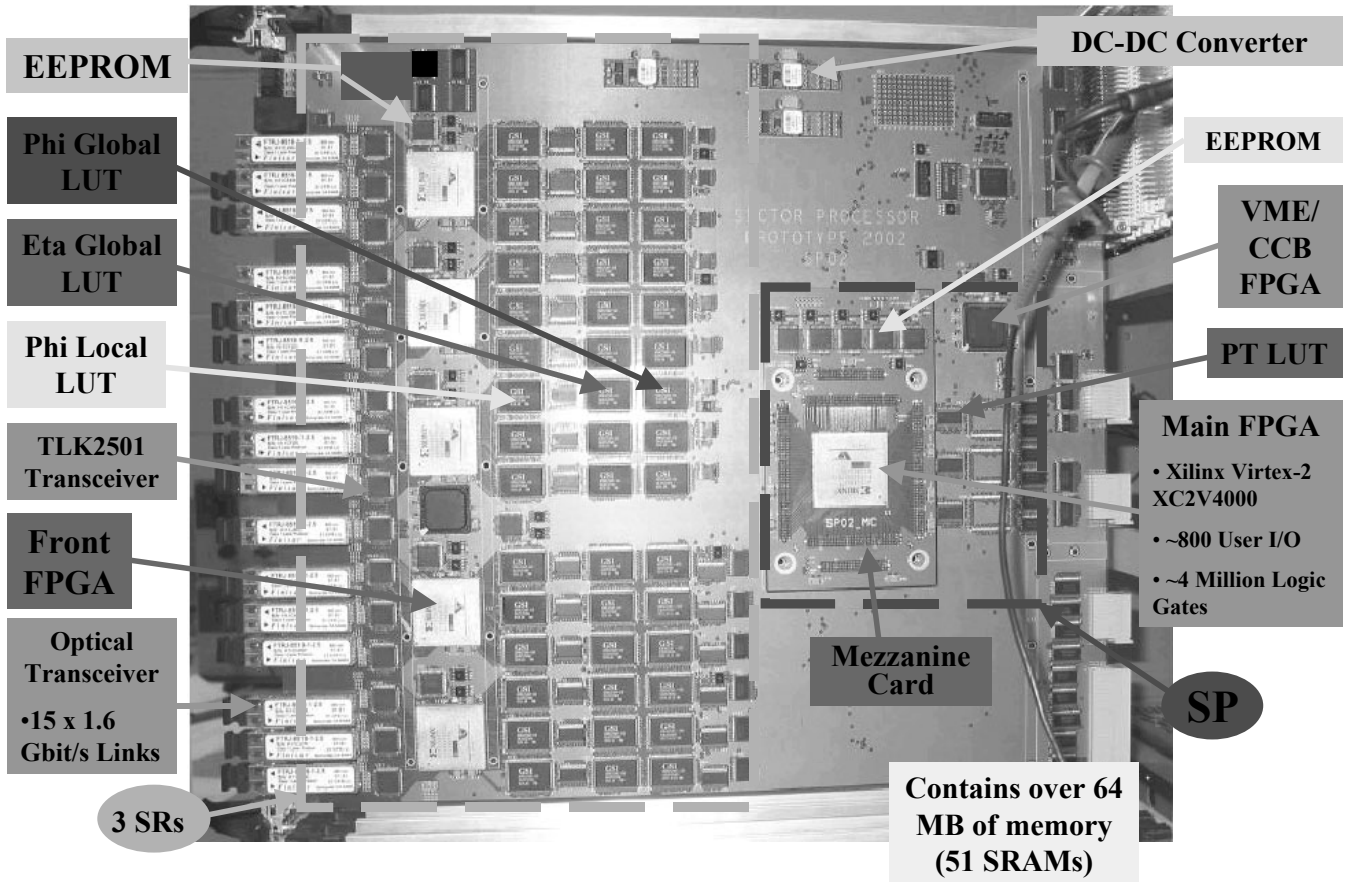

Figure 3: Second Sector Processor prototype

Processor. This FPGA sits on a mezzanine card on the SP, thus allowing for maximum design flexibility for future improvements. In the first prototype this algorithm required 15 FPGAs.

The output of the Sector Processor FPGA is sent to the $P_T$ assignment lookup tables, also GSI SRAMS [7], and the results of the $P_T$ assignment for the three best muons are sent via the custom backplane to the Muon Sorter.

In the second prototype Track-Finder system, we no longer use Channel Links for the backplane transmission because of their long serialization/deserialization latency of 100 ns; rather, GTLP backplane technology is now used. This allows transmitting the data point-to-point (from Sector Processor to Muon Sorter) at 80 MHz, with no time penalty for serialization since the most time-critical portions of data are sent. The SP will also send data to a DAQ readout board over the SP's sixteenth optical link. This board receives the LCT data received by the SP from the MPC, and final results completed by the SP. The entire second prototype Track-Finder system fits into one 9U VME crate (Figure 5).

The second prototype has been designed and built, and is now undergoing tests. So far, the FPGAs have been successfully programmed, and the VME interface and the onboard databus have been validated. Successful tests have also been completed to verify operation of the TLK2501 chips on the SP. These chips have the ability to generate pseudo random bit streams to facilitate testing of a single chip (internal loop-back test), or pairs of chips. In transmitting these bit streams over optical fibres, one TLK2501 is used to send data, and another is

Card and a Sector Processor. The MPC sat in the same VME crate as the SP using a special test slot designed in the custom backplane. Test routines were written which load test LCT patterns into the MPC input buffer, and transmit a subset of these patterns over the optical links into the SP. The output LCTs from the MPC were checked against the SP input LCTs, and were found to be in agreement.

The original design for the optical links called for a common clock for the operation of transceivers and for trigger logic; however, it is enough to have a low-jitter clock (e.g. from a crystal) operating at the transmitter frequency, which act as a reference for the transceivers, and a second clock driving the SP logic; such a design would allow the links to operate more robustly with respect to clock jitter in the distributed LHC clock. More tests will be conducted between the MPC and SP to determine the most optimal clock configuration.

Future tests also include: operation of Sector Receiver memories and verification of Track-Finding logic. There will also be opportunities to examine the functionality of the Sector Processor during a possible used to receive data. Because the SP is equipped with optical transceivers, one can transmit and receive these bit streams on the same SP by utilizing multiple links. Such tests were also completed between a Muon Port future beam test, and with the use of cosmic rays.

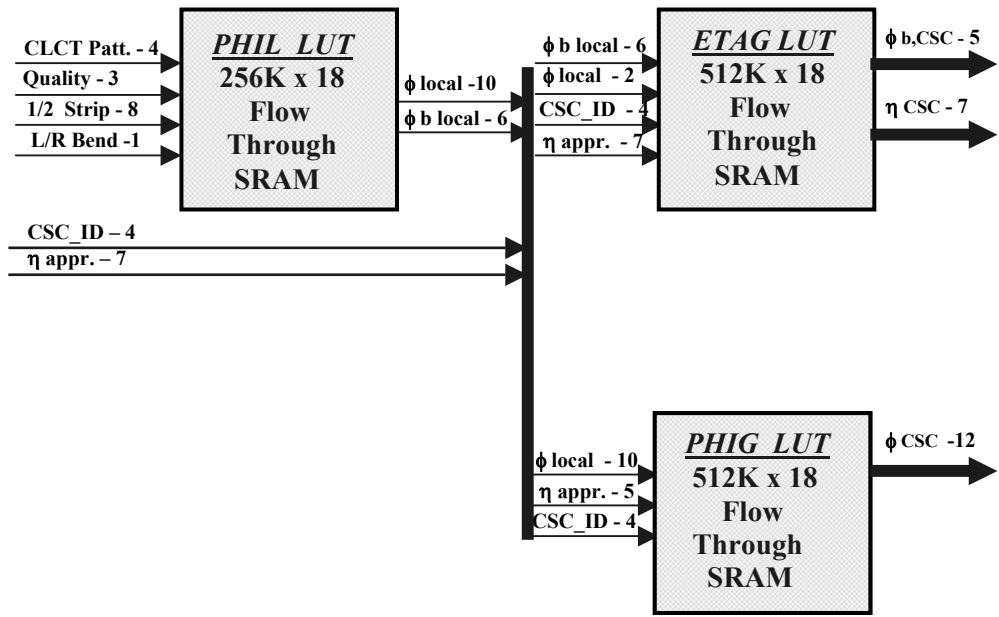

**Figure 4:** Sector Receiver look-up memory scheme. Here "CLCT Patt", "Quality", "½ Strip", "L/R Bend", "CSC_ID", and "η appr." are variables that describe a LCT received by the Sector Receiver. From this data, $\phi$ with respect to the given chamber ("$\phi$ local"), and the bend angle ("$\phi$b local") are found. These are then sent to more memories from which $\phi$ ("$\phi$ CSC-12") and $\eta$ ("$\eta$ CSC-7") are reported.

All software used for testing of the second prototype was written using the Hardware Access Library (HAL), which is part of the XDAQ software package from CMS

DAQ [8]. This library provides a global set of primitive commands independent of the particular bus adapter used for VME communication; thus, maximizing the portability of command sequences generated to control a given VME hardware module. All tests performed on the MPC and SP used special C++ classes developed from the same primitive commands. This allowed easy development of control scripts for two distinct VME modules each with unique firmware control registers.

These SP control classes have been wrapped into special super-classes to be used within the XDAQ environment. This will allow communication of VME modules in multiple VME crates, over a network. With these XDAQ classes, one can control multiple modules, in multiple crates, with the use of a single interface; thus concealing the control register layer offered by the firmware of each hardware module from the user. These tools will ultimately be used for the development of a run-control of the endcap system.

- The algorithms of the extrapolation and final selection units are reworked, and now each of them is completed in only one clock.
- The Track Assembler Units in the first prototype were implemented as external lookup tables (static memory). For the second prototype, they are implemented as FPGA logic. This saved I/O pins on the FPGA and one clock period of latency.
- The preliminary calculations for the $P_T$ assignment are done in parallel with final selection for all 9 muons, so when three best out of nine muons are selected, the pre-calculated values are immediately sent to the external $P_T$ assignment lookup tables.

All this allowed reducing the latency of the Sector Processor algorithm (FPGA plus $P_T$ assignment memory) down to 5 clocks (125 ns) from 15 clocks in the first prototype.

During the construction and debugging of the first prototype, we encountered many problems related to the correspondence between hardware and C++ model. In particular, sometimes it is very problematic to provide

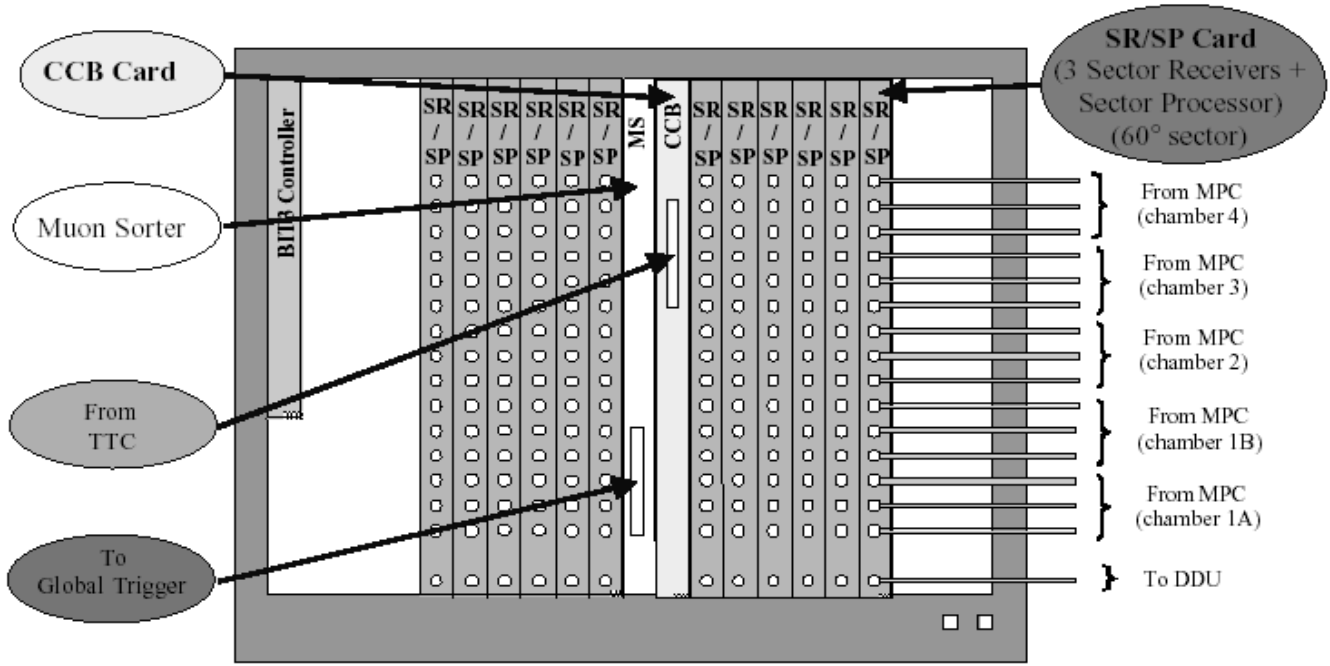

**Figure 5: New crate layout for the Track-Finding system**

## 5. SECTOR PROCESSOR ALGORITHM AND C++ MODEL MODIFICATIONS

The Sector Processor algorithm was significantly modified from that used in the first prototype in order to fit into one chip and reduce latency. In particular, the following modifications were made:

the exact matching, especially if the model uses the C++ built-in library modules, such as lists and list management routines, etc. To eliminate these problems a class library has been developed that allows one to write both the simulation code and firmware in C++, and then translate this code into Verilog HDL. Thus, our code serves a dual purpose: it can be compiled either for simulation purposes, or as the generator of the Verilog code used for FPGA programming. This guarantees a bit-for-bit compatible simulation. This Verilog code can then be synthesized by our FPGA vendor tools and is used as our SP Firmware. This allows us to verify the SP logic through C++ debugging tools such as MS Visual C++. We can also run this code as a part of the CMS simulation and reconstruction framework - thus allowing us to use usual analysis tools for verification (e.g. ROOT); therefore, a line-by-line correspondence is maintained between simulation logic and Firmware logic.

## 6. SUMMARY

The design of a Track-Finder for the Level-1 trigger of the CMS endcap muon system is mature and has been successfully prototyped. The design is implemented as 12 identical processors, which cover the pseudo-rapidity interval $0.9 < \eta < 2.4$. The track-finding algorithms are three-dimensional, which improves the background suppression. The $P_T$ measurement uses data from 3 endcap stations, when available, to improve the resolution to 22%. The input to the Track-Finder can be held for more than one bunch crossing to accommodate timing errors. The latency is expected to be 7 bunch crossings (not including the optical link and timing errors accommodation). The design is implemented using Xilinx Virtex FPGAs and SRAM look-up tables and is fully programmable. The first prototype was successfully built and tested; the pre-production prototype has been built and is currently undergoing tests.

## 7. REFERENCES

[1] CMS Level-1 Trigger Technical Design Report, section 12.4. CERN/LHCC 2000-038

[2] D. Acosta *et al*. "Development and Test of a Prototype Regional Track-Finder for the Level-1 Trigger of the Cathode Strip Chamber Muon System of CMS," NIM A496 (2003) 64

[3] National Semiconductor, DS90CR285/286 datasheet.

[4] D. Acosta *et al*. "The Track-Finding Processor for the Level-1 Trigger of the CMS Endcap Muon System." Proceedings of the LEB 1999 Workshop.

[5] Xilinx Inc., www.xilinx.com

[6] GSI G58320Z18T datasheet

[7] GSI GS881Z18AT datasheet

[8] V. Brigljevic *et al*. "Using XDAQ in Application Scenarios of the CMS Experiment." CHEP, March 2003, La Jolla, California.